\documentclass[10pt]{article}
\usepackage{graphicx}
\usepackage[T1]{fontenc}
\usepackage[utf8]{inputenc}
\usepackage{amssymb}
\usepackage{amsfonts}
\usepackage{dsfont}
\usepackage{mathtools}
\usepackage{amsthm}
\usepackage{amsmath}
\usepackage{relsize}
\usepackage{textcomp}
\usepackage{eurosym}
\usepackage{stmaryrd}
\usepackage{xcolor}
\usepackage[multiple]{footmisc}

\usepackage{geometry}
\begin{document}

\title{Market making by an FX dealer: tiers, pricing ladders and hedging rates for optimal risk control}

\author{Alexander \textsc{Barzykin}\footnote{HSBC, 8 Canada Square, Canary Wharf, London E14 5HQ, United Kingdom, alexander.barzykin@hsbc.com} \and Philippe \textsc{Bergault}\footnote{CMAP, Ecole Polytechnique, Route de Saclay, 91120 Palaiseau, France, philippe.bergault@polytechnique.edu} \and Olivier \textsc{Guéant}\footnote{Centre d'Economie de la Sorbonne, Université Paris 1 Panthéon-Sorbonne, 106 Boulevard de l'Hôpital, 75642 Paris Cedex~13, France, olivier.gueant@univ-paris1.fr}}
\date{}

\maketitle

\begin{abstract}

\vspace{5mm}

Dealers make money by providing liquidity to clients but face flow uncertainty and thus price risk. 
They can efficiently skew their prices and wait for clients to mitigate risk (internalization), or trade with other dealers in the open market to hedge their position and reduce their inventory (externalization). 
Of course, the better control associated with externalization comes with transaction costs and market impact. 
The internalization vs. externalization dilemma has been a topic of recent active discussion within the foreign exchange (FX) community. 
This paper offers an optimal control framework for market making tackling both pricing and hedging, thus answering a question well known to dealers: `to hedge, or not to hedge?'\\

\end{abstract}

\setlength\parindent{0pt}

\section*{Introduction}

With more than \$6 trillion of trading turnover per day, the foreign exchange (FX) market is the largest financial market, far ahead that of bonds and stocks (see \cite{schrimpf2019sizing}). 
In spite of its size and the concentration of trading in a few financial global hubs, FX remains a highly fragmented over-the-counter (OTC) market with, on one side, a dealer-to-client (D2C) segment where dealers / market makers provide liquidity to clients and, on the other, a dealer-to-dealer (D2D) or inter-dealer segment where dealers trade together, mainly for hedging purpose.\\

Market makers in FX cash markets provide liquidity to customers by proposing prices at which they are ready to buy and sell currency pairs through electronic price streams, single-bank platforms, multi-bank platforms, etc. 
As a consequence of the trading flow coming from their clients, they have to manage risky positions. 
They can have two different behaviors: holding the risk until other clients come to offset it (internalization) or hedging the risk out by trading on the D2D segment (externalization). 
Externalization allows to get rid of the risk but it usually comes at a cost, that of crossing the bid-ask spread and sometimes walking the book on platforms like EBS (part of CME Group) or Refinitiv (depending on the currency pair). 
Furthermore, externalization usually induces market impact because trades become visible to more market participants. 
Internalization allows to avoid market impact, or at least reduce it, but this is of course risky for the market maker because price might evolve adversely before the trading flow compensates the current position. The risk can be reduced by skewing prices to attract the flow in the required direction but the flow is by no means guaranteed. In practice, most market makers both internalize and externalize depending on the market conditions (an increase in volatility increases the propensity to externalize) and their positions (dealers usually externalize beyond a certain position limit).\\

The last BIS Triennial Survey has documented the growing prevalence of internalization and the resulting decline of the D2D segment (see \cite{schrimpf2019fx}).\footnote{The report even highlights that electronification has receded on the D2D segment whereas it has grown at fast pace on the D2C segment.} However, the trade-off between internalization and externalization has attracted little academic interest until recently. 
In fact, most of the models proposed in the literature on optimal OTC market making have assumed no way to hedge out the risk through the inter-dealer segment of the market.\footnote{Many market making models in limit order books authorize the use of market orders to unwind part of the risk -- see for instance \cite{guilbaud2013optimal} -- but these models are not adapted to our OTC problem.} In the old paper by Ho and Stoll \cite{ho1981optimal} and in the numerous papers (see for instance \cite{bergault2021size}, \cite{cartea2014buy}, \cite{gueant2017optimal}, \cite{gueant2013dealing}) of the new literature on optimal market making that has followed the paper \cite{avellaneda2008high} of Avellaneda and Stoikov, the market maker is indeed `only' proposing bid and ask quotes.\footnote{Some of these papers did not address specifically OTC markets but the models they proposed are more adapted to OTC markets than stock markets mainly organized around all-to-all limit order books.} The rare references to the internalization vs. externalization dilemma include the paper \cite{bakshaev2020market} by Bakshaev who proposed a reinforcement learning formulation and the paper \cite{butz2019internalisation} by Butz and Oomen who discussed internalization on the basis of queuing theory and derived typical internalization horizons.\\

We have recently proposed in \cite{barzykin2021algorithmic} a model of algorithmic market making with pricing and hedging in a unified stochastic optimal control framework thus constituting an important and natural encounter between two problems that have attracted a lot of academic and practitioners' interest in the last decade: optimal market making and optimal execution.\footnote{For an introduction to optimal execution problems, we refer to the original paper \cite{almgren2001optimal} by Almgren and Chriss and the two reference books \cite{cartea2015algorithmic, gueant2016financial}.} 
The model allows one to set optimal pricing ladder and determine optimal hedging rate in external liquidity pools as functions of the inventory, risk aversion and market-driven parameters. 
In particular, we have proven the existence of a pure flow internalization area, or equivalently, an inventory threshold below which it is optimal for the dealer not to externalize. 
This threshold is derived from a subtle balance between uncertainty, execution costs and market impact.\\

In this paper, we generalize our algorithmic market making model further to include a better modelling of the trading flow. In particular, we introduce tiers in order to account for the complexity of the D2C segment of the market. Tiers are used by market makers to distinguish both the different sources of the trading flow and the natural diversity of the clients. We describe below our modelling approach of the trading flow and show how to estimate the intensity parameters. We demonstrate that tiers can be conveniently defined using clustering techniques on intensity parameters. We then present our algorithmic market making model and look into the difference of optimal strategies for different tiers. By analyzing typical risk neutralization time and internalization ratio derived from the model as functions of the dealer's risk aversion we recover figures consistent with those of \cite{butz2019internalisation} and \cite{schrimpf2019fx}. Finally, we discuss dealer's efficient frontier and comment on the choice of the risk aversion parameter.\\

\section*{Understanding trading flow}

One of the central issues for a dealer is inventory management. Indeed, a dealer must, at all times, decide whether they wish to warehouse the risk while waiting for the arrival of future customer flows or if they wish to hedge part of it by trading on the D2D market. This decision obviously depends on price risk  but also on customer flow. Furthermore, when the market maker decides to hold risk (internalization), they skew prices in order to increase or decrease the flow of buying or selling customers -- depending on the sign of the inventory. Understanding trading flow and customers' sensitivity to streamed prices is therefore essential.\\

In order to model customer flow as a function of streamed prices, let us introduce first a reference price process $(S_t)_t$. 
Following industry standard, we take the firm primary mid price as the reference price at any point in time (EBS for our examples with EURUSD).\footnote{The notion of reference price can be obscure in the case of FX due to significant geographical delocalization of liquidity and last look practice (see \cite{cartea2019foreign} and \cite{oomen2017lastlook}). The true market price can only be known with limited accuracy. Nevertheless, the primary venue provides a reliable measurable reference, suitable for the purpose of this analysis.}
Given a streamed pricing ladder at the bid (resp. ask/offer) modelled by $S^b(t,z) = S_t(1 - \delta^b(t,z))$ (resp. $S^a(t,z) = S_t(1 + \delta^a(t,z))$) where $z>0$ is the size of the trade,\footnote{Throughout, we shall designate $\delta^{b/a}$ by the term quote although it is only a mark-up or a discount with respect to the reference price.} we assume that buy (resp. sell) trades\footnote{We take the dealer's viewpoint when it comes to trade sides.} of size in $[z, z+dz]$ arrive over the infinitesimal interval $[t, t+dt]$ with probability $\Lambda^b(z,\delta^b(t,z)) dz dt$ (resp. $\Lambda^a(z,\delta^a(t,z)) dz dt$). In practice, dealers propose pricing ladders for a set $\{z_k, 1\le k \le K\}$ of sizes. In what follows, $K=6$ sizes are considered, corresponding to $1$, $2$, $5$, $10$, $20$ and $50$ M\euro\ respectively. As a consequence, the measures $\Lambda^{b/a}(z,\delta)dz$ are approximated by discrete measures $\sum_{k=1}^K\Lambda^{b/a}_k(\delta) 1_{z_k}(dz)$ (where $1_{z_k}(dz)$ is the Dirac measure in $z_k$). Hereafter, the functions $\Lambda_k^{b/a}$ are called intensity functions or simply intensities.\\

For an anonymized sample of HSBC FX streaming clients\footnote{This sample is sufficiently diverse to provide realistic results, but by no means complete to fully represent HSBC FX market making franchise.} trading EURUSD we got access to tables of trades and quotes over the period from January to April 2021.\footnote{In what follows, we focused only on the most liquid hours from 6 a.m. to 8 p.m. UTC.} For the purpose of our study, quotes on the bid and ask sides can be summed up, for each size $z_k$, by a list of couples $((\delta_j, \tau_j))_{j \in \mathcal J^{b/a}_k}$  where $\delta_j$ is a streamed quote for size $z_k$ and $\tau_j$ the associated duration of that quote. Trades are not all of size in $\{z_k, 1\le k \le K\}$ but we can associate each trade with the closest $z_k$ and the trade data can then be summed up, for the bid and ask sides and for each size $z_k$, by a list of quotes $(\delta_i)_{i \in \mathcal I^{b/a}_k}$.\\

It is easy to show that the log-likelihoods $\textrm{LL}^{b/a}_k$ associated with the bid and ask sides for size $z_k$ are (up to an additive constant):
\begin{eqnarray*}
\textrm{LL}^{b/a}_k &=& \sum_{i \in \mathcal I^{b/a}_k} \log(\Lambda^{b/a}_k(\delta^i)) - \sum_{j \in \mathcal J^{b/a}_k} \Lambda^{b/a}_k(\delta^j)\tau^j\\
&=& I^{b/a}_k \int_{\delta} \log(\Lambda^{b/a}_k(\delta)) f^{b/a,\mathcal{T}}_{k}(d\delta) - \bar{\tau} \int_{\delta} \Lambda^{b/a}_k(\delta) f^{b/a,\mathcal{Q}}_{k}(d\delta)
\end{eqnarray*}
where $I^{b/a}_k$ is the number of trades bucketed with size $z_k$, $f^{b/a,\mathcal{T}}_{k}(d\delta)$ is the probability measure of bid/ask trades bucketed with size $z_k$, $f^{b/a,\mathcal{Q}}_{k}(d\delta)$ is the probability measure of streamed quotes (weighted with durations) at the bid/ask for size $z_k$ and $\bar\tau = \sum_{j \in \mathcal J^{b/a}_k} \tau^j$ is the total duration of the time window.\\

Intensity functions can be interpreted in the following two-step fashion. First, there is a given flow of prospective customers who look at the prices. The probability that they trade depends then on the quotes proposed by the dealer. Therefore, inspired by logistic regression techniques, a natural functional form is $\Lambda^{b/a}_k(\delta) =  \frac{\lambda^{b/a}_k} {1 + e^{\alpha^{b/a}_{k} + \beta^{b/a}_k \delta}}$, where $\lambda^{b/a}_k$ represents the flow of prospective customers and the term $\frac{1} {1 + e^{\alpha^{b/a}_{k} + \beta^{b/a}_k \delta}}$ represents the probability of trading given the quotes proposed. By using a maximum likelihood approach, one can easily estimate the parameters, i.e. for each $k$,
$$(\lambda^{b/a}_k,\alpha^{b/a}_k, \beta^{b/a}_k) \in \textrm{argmax} \left(I^{b/a}_k \int_{\delta} \log\left(\frac{\lambda^{b/a}_k} {1 + e^{\alpha^{b/a}_{k} + \beta^{b/a}_k \delta}}\right) f^{b/a,\mathcal{T}}_{k}(d\delta) - \bar{\tau} \int_{\delta} \frac{\lambda^{b/a}_k} {1 + e^{\alpha^{b/a}_{k} + \beta^{b/a}_k \delta}} f^{b/a,\mathcal{Q}}_{k}(d\delta)\right).$$

While carrying out the above estimation procedure on individual clients, we noticed that intensities on the bid and ask sides were not significantly different. Therefore, we assumed $\Lambda_k^b(\delta) = \Lambda_k^a(\delta) = \Lambda_k(\delta)$. This assumption enabled to have more precise estimations since, then, bid and ask tables could be concatenated and the log-likelihoods added. For the examples of this paper, we therefore fit the functions $\Lambda_k(\delta) =  \frac{\lambda_k} {1 + e^{\alpha_{k} + \beta_k \delta}}$, by choosing, for each $k$,
$$(\lambda_k,\alpha_k, \beta_k) \in \textrm{argmax} \left( I_k \int_{\delta} \log\left(\frac{\lambda_k} {1 + e^{\alpha_{k} + \beta_k \delta}}\right) f^{\mathcal T}_{k}(d\delta) - 2\bar{\tau} \int_{\delta} \frac{\lambda_k} {1 + e^{\alpha_{k} + \beta_k \delta}} f^{\mathcal Q}_{k}(d\delta)\right)$$
where $ I_k = I_k^b + I_k^a$, $f^{\mathcal T}_{k} = \frac{I_k^b f^{b,\mathcal{T}}_{k} + I_k^a f^{a,\mathcal{T}}_k}{I_k^b + I_k^a}$ and $f^{\mathcal{Q}}_{k} = \frac{f^{b,\mathcal{Q}}_k + f^{a,\mathcal{Q}}_{k}}2$.\\
 
\begin{figure}[h!]\centering
\includegraphics[width=\textwidth]{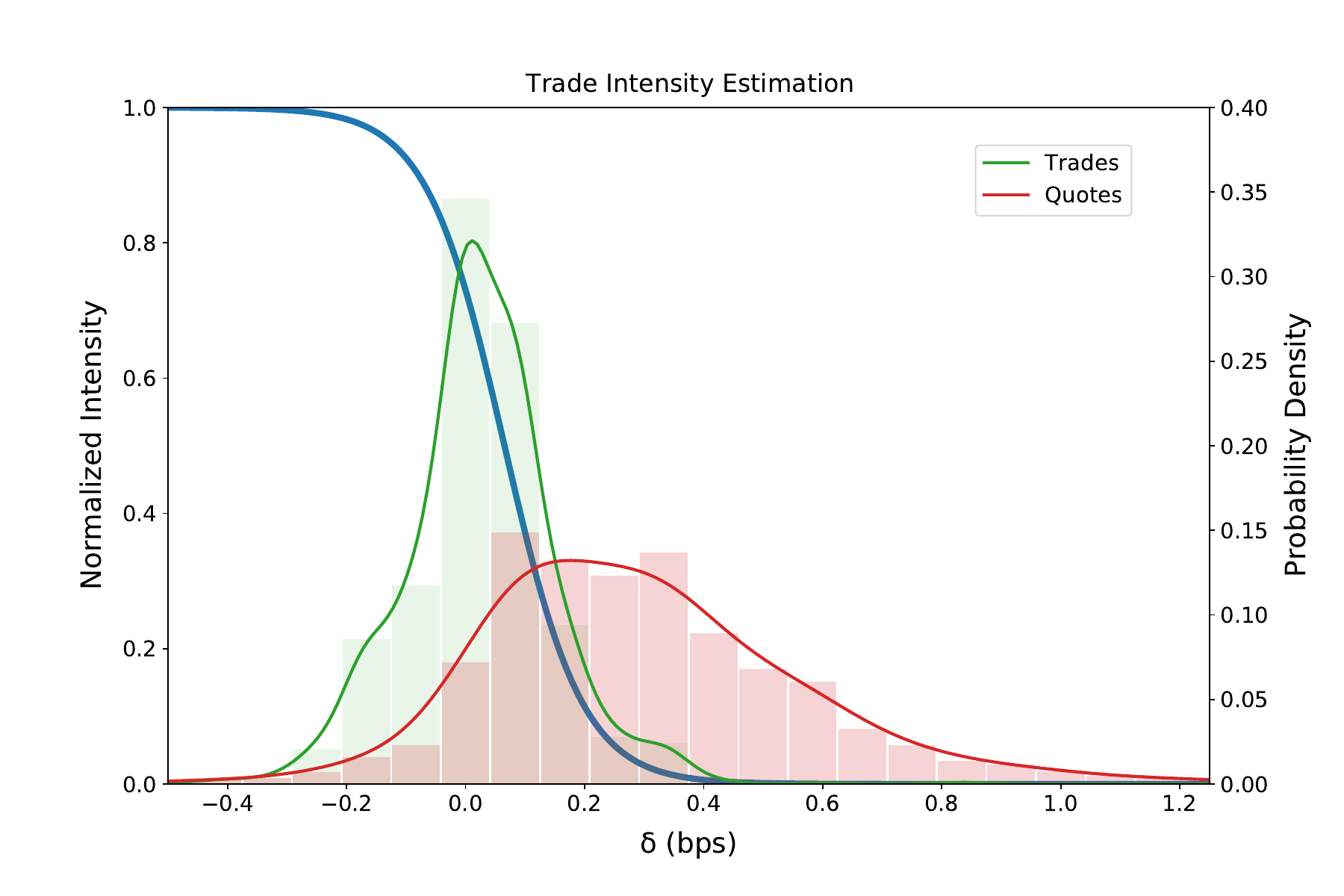}\\
\caption{
Trade (green) and streamed quote (red) frequency histograms and smoothed probability density functions (associated with $f^{\mathcal{T}}_{1}(d\delta)$ and $f^{\mathcal{Q}}_{1}(d\delta)$) for a client chosen at random in our sample and trades of 1~M\euro\ (right axis) along with the corresponding estimated intensity function
(blue, left axis) (normalized to a maximum of $1$). Smooth distributions were obtained using kernel density estimation.}
\label{intensity_estimation}
\end{figure}

The results for $z_1 = 1$ M\euro\  (corresponding to trades of 1~M\euro) are shown in Fig.~\ref{intensity_estimation} for a single client chosen at random in our sample. We do not document the scale (i.e. $\lambda_1$) but only the shape (corresponding to the function $\delta \mapsto \frac{1}{1+e^{\alpha_1 + \beta_1 \delta}}$).\\

\begin{figure}[h!]\centering
\includegraphics[width=0.95\textwidth]{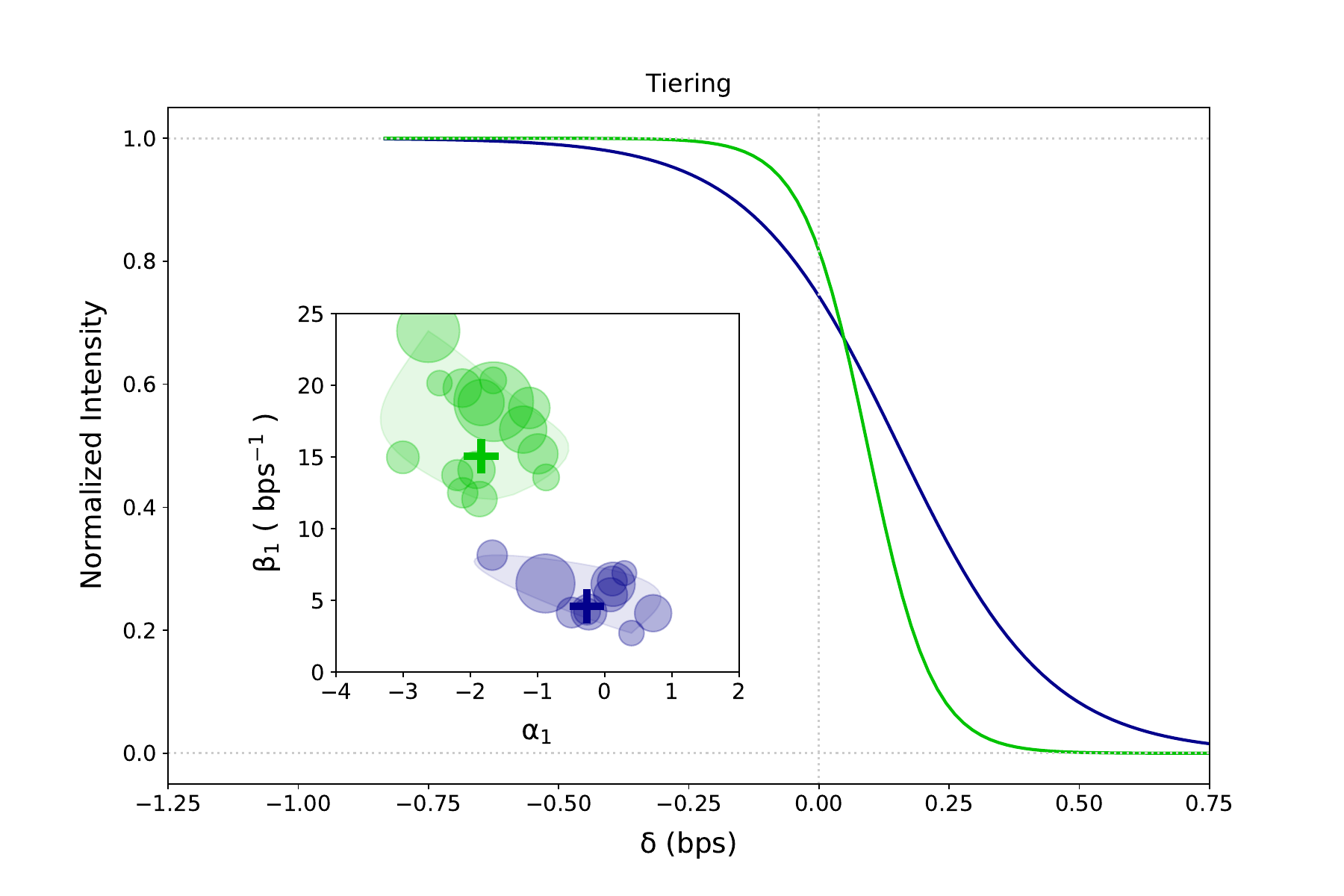}\\
\caption{
Normalized intensity functions for the two client tiers (Tier 1 is in blue and Tier 2 in green). Tiers are identified using standard $k$-means procedure on individually fit intensity parameters 
for the sample of clients considered in the paper and trades of 1~M\euro\ (insert).}
\label{tiering}
\end{figure}

Fig.~\ref{tiering} collects $(\alpha_1, \beta_1)$ parameters for all clients in the sample. Two clusters are clearly visible, justifying the creation of tiers. The intensity functions corresponding to the two tiers (estimated on pooled data for each tier using the same maximum likelihood approach as above) are shown as well. One can observe significantly different price sensitivities across the two tiers. The respective parameters are (after rounding) $\alpha^1_1 = -0.3$ and $\beta^1_1 = 5$ bps$^{-1}$ for Tier 1 and $\alpha^2_1 = -1.9$ and $\beta^2_1 = 15$ bps$^{-1}$ for Tier 2.
This correlates well with recent findings on informativeness and trading behavior of typical FX OTC market participants, with different pricing sensitivity of different types of clients driven by significantly different business horizons, risk management requirements and information access \cite{oomen2017aggregator, ranaldo2021}.\\ 

For other sizes, shape parameters and therefore tiers were found consistent with the results for trades of 1~M\euro. $(\lambda_1, \ldots, \lambda_6)$ have been rounded and found proportional to $(0.4, 0.25, 0.19, 0.1, 0.05, 0.01)$ for both tiers. We therefore define throughout the paper the parameters $\alpha^1 = -0.3$ and $\beta^1 = 5$ bps$^{-1}$ for Tier 1 and $\alpha^2 = -1.9$ and $\beta^2 = 15$ bps$^{-1}$ for Tier 2.\\

\section*{Market making model for multiple tiers}

Let us now come to the market making model. To be general, we denote by $N$ the number of tiers ($N=2$ in our examples). The market maker streams a pricing ladder for each tier: to tier $n \in \{1,\ldots, N\}$ they propose a pricing ladder $S^{b,n}(t,z) = S_t(1 -\delta^{b,n}(t,z))$ at the bid and $S^{a,n}(t,z) = S_t(1 +\delta^{a,n}(t,z))$ at the ask. The associated intensities for tier $n$ are denoted by $\Lambda^{b,n}$ and $\Lambda^{a,n}$, respectively. Following the above empirical results, we assume that the functions $\Lambda^{b,n}$ and $\Lambda^{a,n}$ have the form\footnote{Generalizations are of course straightforward.}
$$\Lambda^{b,n}(z,\delta)=\Lambda^{a,n}(z,\delta) = \Lambda^n (z,\delta) = \lambda^n(z) f^n(\delta) \quad \text{with} \quad f^n(\delta) =  \frac{1}{1+e^{\alpha^n + \beta^n \delta}}.$$

The market maker can also trade on a platform to hedge their position. The execution rate of the market maker on this platform is modelled by a process $(v_t)_t$.\\

Coming to the dynamics of the reference price, we assume it has two parts: an exogenous part with classical lognormal dynamics and an endogenous part corresponding to the permanent market impact of the market maker's trades on the platform (i.e. when they externalize). Mathematically, $(S_t)_t$ has the dynamics
$$dS_t = \sigma S_t dW_t + k v_t S_t dt,$$
where $(W_t)_t$ is a standard Brownian motion, $k$ represents the magnitude of the (linear) permanent impact, and $\sigma$ is the volatility parameter.\\

We denote by $(q_t)_t$ the inventory process of the market maker. $q_t$ stands therefore for the position at time $t$ of the market maker resulting from trades with clients and trades on the platform. Mathematically, denoting by $J^{b,n}(dt,dz)$ and $J^{a,n}(dt,dz)$ the random measures modelling the times and sizes of trades with tier $n$ on the bid and ask sides, respectively, the dynamics of $(q_t)_t$ is given by 
$$dq_t = \sum_{n=1}^N \int\limits_{z=0}^{\infty} zJ^{b,n}(dt,dz) - \sum_{n=1}^N \int\limits_{z=0}^{\infty} zJ^{a,n}(dt,dz) + v_t dt.$$

The resulting cash process $(X_t)_t$ of the market maker writes 
$$dX_t =  \sum_{n=1}^N \int\limits_{z=0}^{\infty} S^{a,n}(t,z)zJ^{a,n}(dt,dz) -  \sum_{n=1}^N \int\limits_{z=0}^{\infty} S^{b,n}(t,z)zJ^{b,n}(dt,dz) - v_tS_t dt - L(v_t) S_tdt,$$
where the term $L(v_t)S_t$ accounts for the execution costs when they externalize.\footnote{$L$ is typically nonnegative, strictly convex, and asymptotically super-linear.}\\

The market maker wants to maximize the expected Mark-to-Market value of their portfolio at the end of the period $[0,T]$, while managing the risk associated with their inventory. Mathematically, we assume that they want to maximize
$$\mathbb E \left[ X_T + q_T S_T - \frac{C}{2} \int_0^T q_t^2 d[S]_t \right]$$ by choosing $\delta^{b,n}$, $\delta^{a,n}$, and $v$, where the respective importance of the expected P\&L and risk management components can be chosen through the coefficient $C \ge 0$.
This is a standard objective function discussed in the market making literature (see \cite{gueant2016financial}).\footnote{
One can introduce a terminal penalty on the residual inventory at time $T$ but this does not change the stationary quotes we obtain even for short-term time horizon $T$ in the case of FX.}
Market share is also often targeted by dealers, and this could be part of a more general multi-objective optimization problem, but P\&L and risk would always remain at core.\\

Applying Itô's formula to the process $\left(X_{t} + q_t S_t  \right)_{t}$ allows to see that this problem is equivalent to maximizing
\begin{equation*}
\mathbb{E}\left[\int\limits_{0}^{T} \Bigg(  \sum_{n=1}^N \int\limits_{0}^{\infty}\Big(z\delta^{b,n}(t,z)  \Lambda^{n}(\delta^{b,n}(t,z))\! +\! z \delta^{a,n}(t,z)  \Lambda^{n}(\delta^{a,n}(t,z)) \Big)S_tdz\! +\! k q_t v_t S_t \!-\! L(v_t)S_t\! - \! \frac{C}{2} \sigma^2q_{t}^2 S_t^2 \Bigg) dt  \right].
\end{equation*}

As $T$ is chosen small in what follows, it makes sense to approximate $S_t$ by $S_0$ in the above and the problem becomes that of maximizing
\begin{equation*}
\mathbb{E}\left[\int\limits_{0}^{T} \Bigg(  \sum_{n=1}^N \int\limits_{0}^{\infty}\Big(z\delta^{b,n}(t,z)  \Lambda^{n}(\delta^{b,n}(t,z)) + z \delta^{a,n}(t,z)  \Lambda^{n}(\delta^{a,n}(t,z)) \Big)dz + k q_t v_t - L(v_t) -  \frac{\gamma}{2} \sigma^2q_{t}^2 \Bigg) dt  \right],
\end{equation*}
where $\gamma = C S_0$ is analogous to the risk aversion parameter in most models of the market making literature.\\

We denote by $\theta:[0,T]\times \mathbb R\rightarrow \mathbb{R}$ the value function of this stochastic control problem. The Hamilton-Jacobi equation associated with it is:
\begin{eqnarray*}
0 &=& \partial_t \theta(t,q) - \frac{\gamma}{2}\sigma^2 q^2 + \sum_{n=1}^N \int\limits_{0}^{\infty} zH^{n} \left(\frac{\theta(t,q) -  \theta(t,q+z) }{z}\right)\lambda^n(z) dz\\
&& + \sum_{n=1}^N \int\limits_{0}^{\infty} zH^{n} \left(\frac{\theta(t,q) - \theta(t,q-z)}{z} \right)\lambda^n(z) dz +  \mathcal H \left(\partial_{q}\theta(t,q) +kq \right),\qquad  \forall (t,q) \in [0,T)\times\mathbb{R}  
\end{eqnarray*}
with terminal condition $\theta(T,\cdot) = 0$,
where
$$
H^{n}:p\in\mathbb{R} \mapsto \underset{\delta }{\sup} f^n(\delta)(\delta-p) \qquad \textrm{and} \qquad 
    \mathcal H:p\in\mathbb{R} \ \mapsto \underset{v }{\sup} \left(vp- L(v)   \right).$$

Under mild assumptions (see \cite{bergault2021size, gueant2017optimal} for similar results), it can be proved that, given a smooth solution to the above Hamilton-Jacobi equation, the optimal controls are given by\footnote{$(q_{s})_s$ is a càdlàg process and we write $q_{t-} = \lim_{s \uparrow t} q_s$ the left limit of the process $q$ at time $t$.}
$$\delta^{b,n*}(t,z) = \bar\delta^{n} \left( \frac{\theta(t,q_{t-}) -  \theta(t,q_{t-}+z) }{z} \right),$$
$$\delta^{a,n*}(t,z) = \bar \delta^{n} \left( \frac{\theta(t,q_{t-}) -  \theta(t,q_{t-}-z) }{z} \right),$$
and
$$v^*_t = \mathcal H'\left( \partial_{q}\theta(t,q_{t-}) +kq_{t-} \right),$$
where $\bar \delta^{n}(p) = (f^n)^{-1} \left(-{H^n}' (p)  \right)$.

\section*{Numerical results and discussion}

To illustrate the optimal market making strategy in our model, let us focus on the case of a typical top-tier bank dealer on EURUSD. Regarding size buckets, client tiering, and the shape of intensities we used the same parameters as in the above study of the trading flow on a sample of HSBC clients. Regarding intensity amplitudes, we set $(\lambda_1, \ldots, \lambda_6) = \lambda \cdot (0.4, 0.25, 0.19, 0.1, 0.05, 0.01)$ for both tiers with $\lambda = 1800$ day$^{-1}$.
This figure was chosen so that, using the optimal strategy, the trading flow be of the same order of magnitude as the estimation proposed by Butz and Oomen in \cite{butz2019internalisation} (see also BIS data~\cite{schrimpf2019fx}). We obtained approximately 10 billion euros of daily turnover (their estimation was $7.32$ M\$/min for a top-tier bank dealer on EURUSD to be precise).\footnote{Note that the maximum daily turnover corresponding to these parameters is approximately $31$ billion euros. However, the dealer can only hypothetically reach this level by quoting far better prices than the mid price and loosing money.}\\

As far as execution cost and market impact parameters are concerned, we used the procedure outlined in~\cite{almgren2005optimal} on a sample of HSBC execution data and chose (after rounding):
\begin{itemize}
    \item $L : v\in \mathbb R \mapsto \eta v^2 + \phi |v|$ with $\eta = 10^{-5}\ \textrm{bps}\cdot \textrm{day} \cdot (\textrm{M\euro})^{-1}$ and $\phi = 0.1\ \textrm{bps}$.
    \item Permanent market impact: $k=5\cdot 10^{-3}\ \textrm{bps} \cdot (\textrm{M\euro})^{-1}$.
\end{itemize}
We set volatility to $\sigma = 50\ \textrm{bps} \cdot \textrm{day}^{-\frac{1}{2}}$ and considered a time horizon $T = 0.05\ \textrm{days}$ ($72$ minutes) that ensures convergence towards stationary quotes and hedging rate at time $t=0$ (see more on convergence in~\cite{barzykin2021algorithmic}).\\

In order to approximate the value function $\theta$, we added some form of boundary conditions by imposing that no trade that would result in an inventory $|q|  > 250\ \textrm{M\euro}$ is admitted, and used a monotone implicit Euler scheme on a grid with 501 points for the inventory.\\

\begin{figure}[h!]\centering
\includegraphics[width=\textwidth]{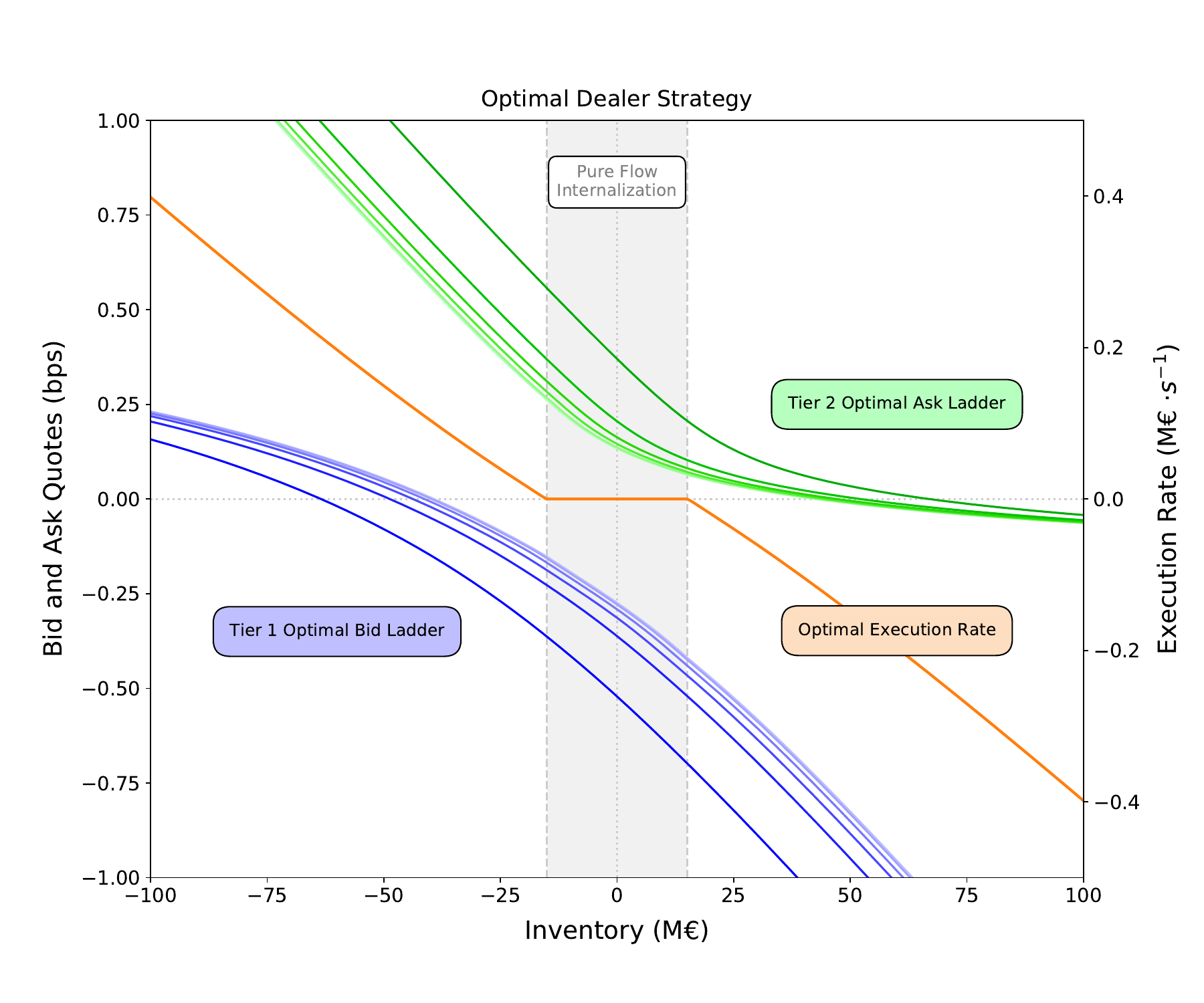}\\
\caption{
Optimal bid ladder for Tier 1 (blue): $-\delta^{b,1*}(0,z)$ as a function of $q_{0-}$ for $z \in \{z_1, \ldots, z_6\}$. Optimal ask ladder for Tier 2 (green): $\delta^{a,2*}(0,z)$ as a function of $q_{0-}$ for $z \in \{z_1, \ldots, z_6\}$. Optimal external hedging rate (orange): $v^*_0$ as a function of $q_{0-}$. Risk aversion parameter: $\gamma=2\cdot 10^{-3}\ \textrm{bps}^{-1} \cdot (\textrm{M\euro})^{-1}$.
}
\label{pricing_hedging}
\end{figure}

\begin{figure}[h!]\centering
\includegraphics[width=\textwidth]{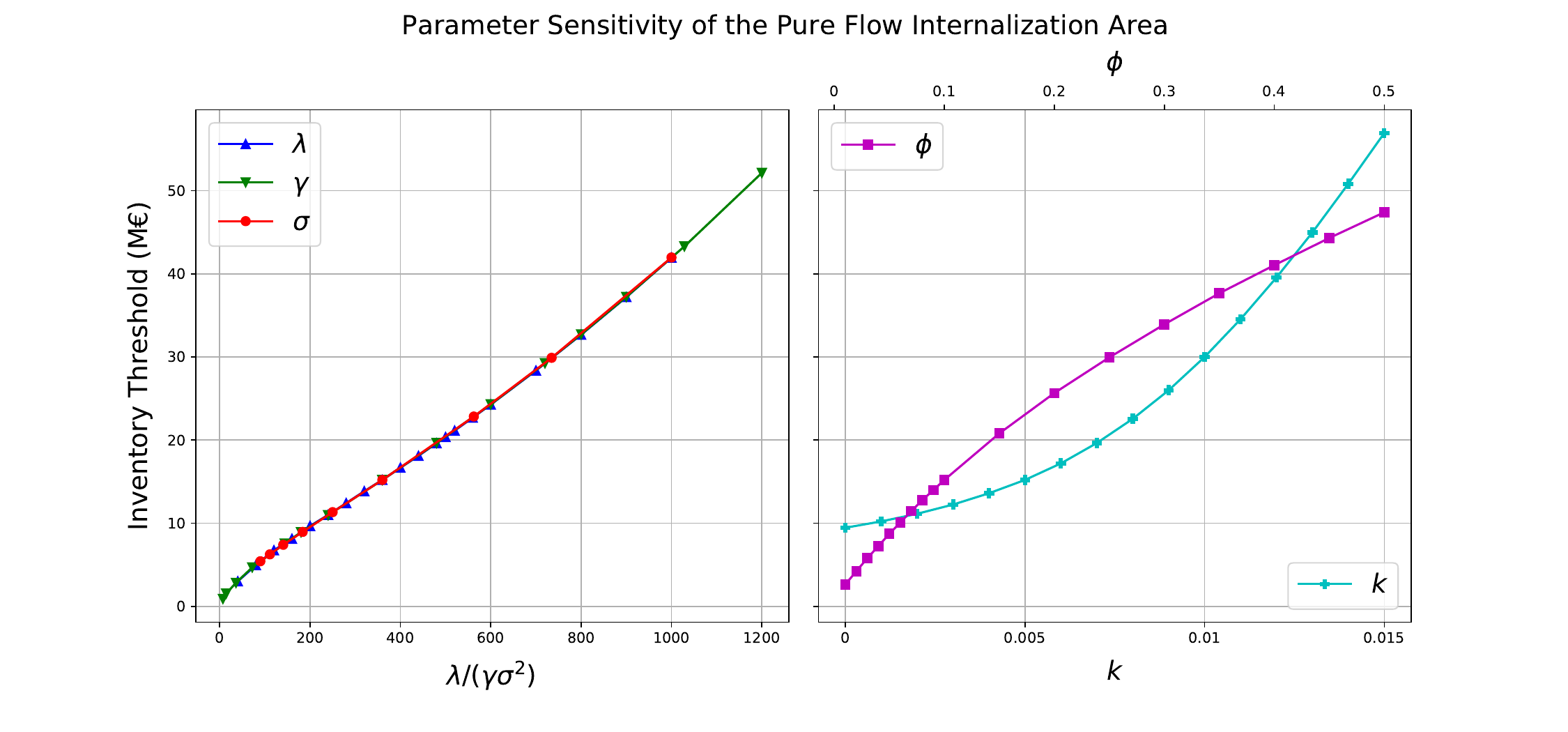}\\
\caption{Inventory threshold of the pure flow internalization area for different levels of risk aversion ($\gamma$), volatility~($\sigma$), franchise size ($\lambda$), execution costs ($\phi$) and permanent market impact ($k$).
Left panel scales against $\lambda/(\gamma \sigma^2)$ with varying parameter colour coded while others being fixed at default values.}
\label{inventory_threshold}
\end{figure}

Fig.~\ref{pricing_hedging} summarizes optimal pricing and hedging strategies for the above set of parameters and risk aversion of  $\gamma=2\cdot 10^{-3}\ \textrm{bps}^{-1} \cdot (\textrm{M\euro})^{-1}$.\footnote{Due to the assumption of flow symmetry, it suffices to plot only bid or ask quotes for each tier as the other side would be a mirror image. We decided to plot $-\delta^{b,1*}(0,z)$ as a function of $q_{0-}$ for $z \in \{z_1, \ldots, z_6\}$ for Tier 1 and $\delta^{a,2*}(0,z)$ as a function of $q_{0-}$ for $z \in \{z_1, \ldots, z_6\}$ for Tier 2.} There are several interesting features deserving to be noticed. Firstly, we observe a range of inventory around zero where the dealer will only internalize by skewing the quotes, i.e. no hedging. We call this interval the pure flow internalization area. Since $L(v) = \eta v^2 + \phi |v|$ implies that $\mathcal H'(p) = \frac1{2\eta}\text{sign}(p) \max(0,|p|-\phi)$, we have $\mathcal H'(p) = 0 \iff |p| \le \phi$. Given the expression of the optimal controls, the pure flow internalization area corresponds to the set of inventories $q$ verifying $\left|\partial_q\theta(0,q) + kq\right| \le \phi$ which contains $0$ plus an interval around $0$ (as soon as $\theta(0,\cdot)$ is continuously differentiable). In terms of sensitivity to the parameters, we empirically noticed (in line with intuition) a wider pure flow internalization area for a less risk averse market maker with a larger franchise, exposed to higher execution costs and market impact and in a less volatile market (see Fig.~\ref{inventory_threshold}). We also note that the optimal execution rate curve is almost linear with respect to inventory outside of the pure flow internalization area.
Secondly, the bid-ask spread is driven by the flow signature, leading to different pricing strategies for the two tiers we considered. Our estimation of inventory-neutral top-of-book bid-ask spread (i.e. the difference between the ask and bid prices proposed by the market maker for notional of 1~M\euro) for price-sensitive clients is 0.26 bps, while for less sensitive clients it is 0.55 bps. This compares well with average composite\footnote{Aggregated order book of multiple ECNs was used.} bid-ask spread of 0.23 bps  and average primary venue bid-ask spread of 0.65 bps at New York open as of this writing (early July 2021). This is particularly interesting as no market bid-ask spread was introduced into the model. We note that market maker's OTC spread is mainly driven by the empirical shape of intensity function.\\

\begin{figure}[h!]\centering
\includegraphics[width=0.95\textwidth]{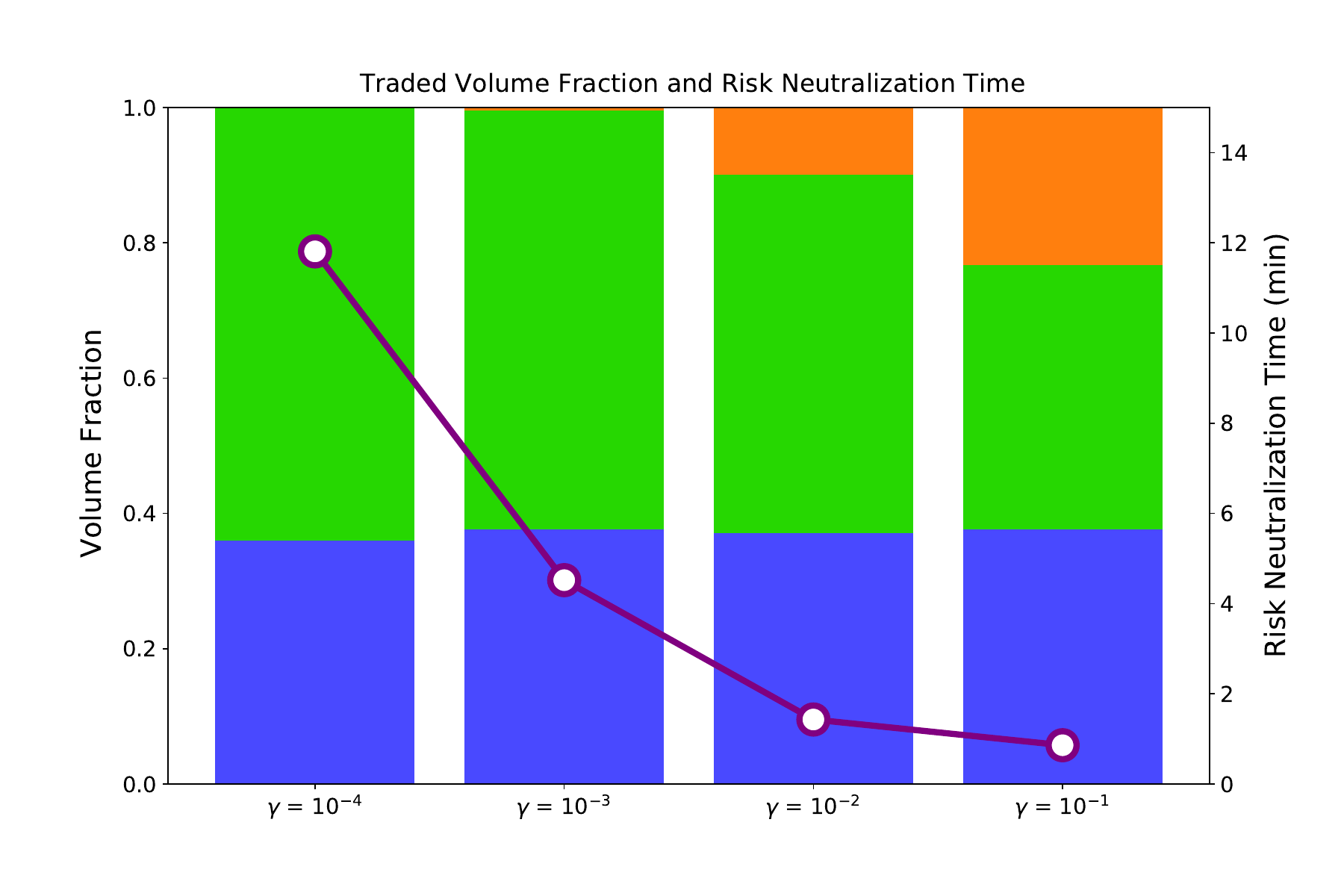}\\
\caption{
Traded volume fraction executed with Tier 1 and Tier 2 clients and externally for hedging purpose (bars, Tier 1 is in blue, Tier 2 in green, and external trading in orange). Risk neutralization time (line and dots) for different values of $\gamma$. 
Results were obtained by Monte Carlo simulation of $10^5$ trajectories over a time horizon $T = 10$ days for a market maker following the stationary optimal strategy.
}
\label{internalization}
\end{figure}

Once the optimal strategy has been computed, one can simulate the behavior of our market maker and assess the volume share of external hedging as compared to different client tiers, for different levels of dealer's risk aversion.
Fig.~\ref{internalization} shows a span of 4 orders of magnitude in $\gamma$ illustrating the crossover from pure internalization to significant externalization.
Note that the volume share of less price-sensitive clients (Tier 1) remains basically the same while the dealer will prefer to sacrifice price-sensitive flow (Tier 2) for the certainty of inventory management when risk aversion increases.
The level of internalization for a risk-aware dealer is in line with BIS reporting around 80\% internalization in G10 currencies by top-tier banks.\\

Fig.~\ref{internalization} also illustrates the dependence of the characteristic risk neutralization time $\tau_R$ on $\gamma$, where $\tau_R$ is defined as the integral of the inventory autocorrelation function. It appears clearly that pure internalization comes with a significantly higher risk.
It is noteworthy that the value of $\tau_R$ for $\gamma=0.01$ compares very well with the EURUSD internalization time estimated by Butz and Oomen (1.39 minutes) in \cite{butz2019internalisation}.\\

\begin{figure}[h!]\centering
\includegraphics[width=0.93\textwidth]{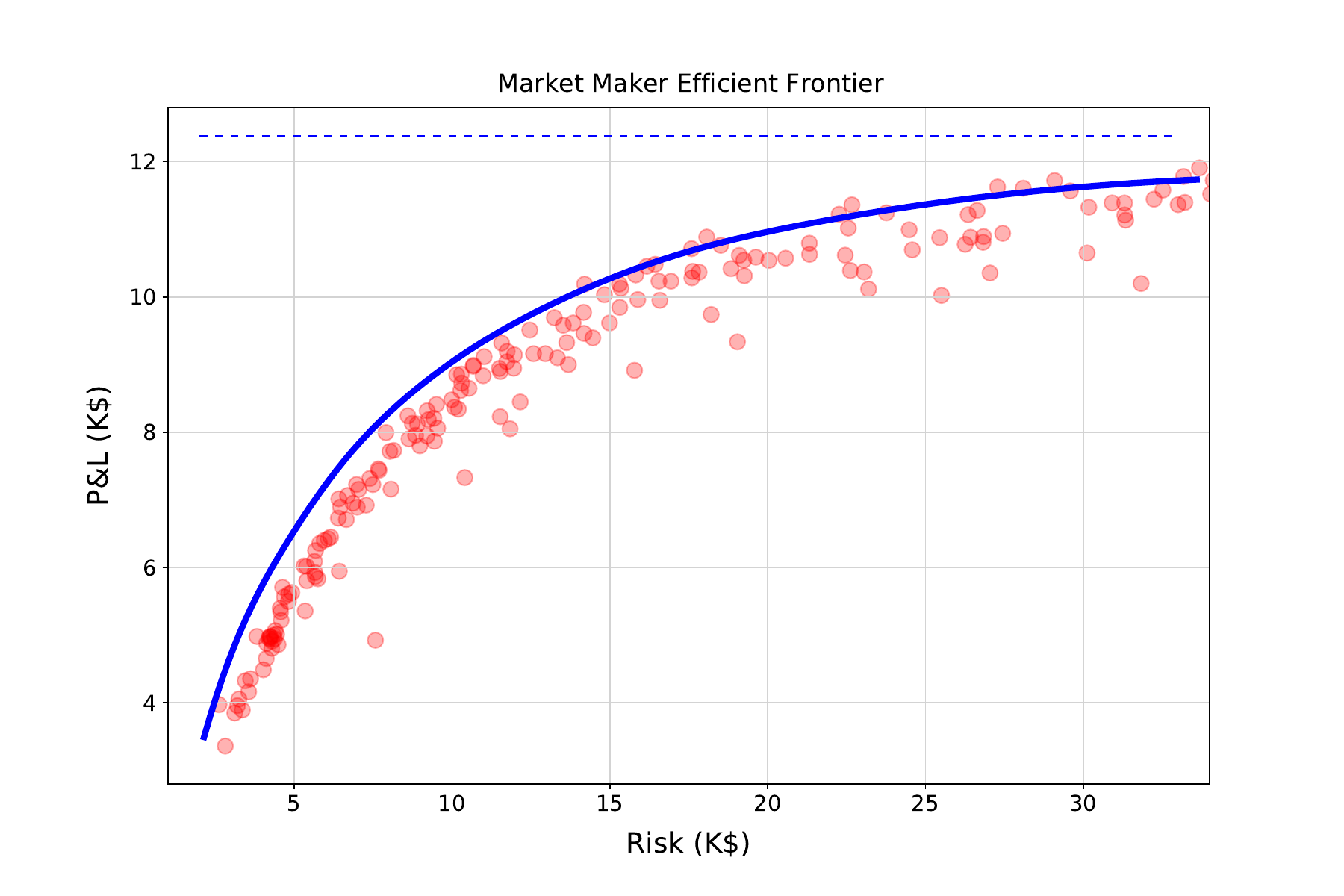}\\
\caption{
Expected P\&L vs. standard deviation of the P\&L over time horizon $T = 0.05$ of a market maker following the stationary optimal strategy with different values of the risk aversion parameter (solid line) and 20 randomly perturbed strategies for each value of $\gamma$ (circles).
Maximum expected P\&L without risk management (dashed line).
Results were obtained by Monte Carlo simulation of $10^5$ trajectories for several values of $\gamma$ ranging from $10^{-4}$ to $10^{-1}$. The curve has been obtained with cubic splines.
}
\label{risk_reward}
\end{figure}

Fig.~\ref{risk_reward} explores the optimal risk-reward trade-off. In order to get the dealer's efficient frontier by analogy with Markowitz modern portfolio theory, we chose different values of the risk aversion parameter $\gamma$ and perturbed the optimal strategy by randomly shifting bid and ask quotes for both tiers and randomly choosing the width of the pure internalization area and the slope of the hedging rate curve around their optimal values.\\
\vspace{-3mm}

The resulting outcomes are almost all below the curve built using the optimal stationary pricing and hedging strategy (smoothed with cubic splines) although our objective function is not exactly a mean-variance one. Our penalty for inventory risk ignores indeed part of the variance (see the discussion on objective functions for market making in \cite{gueant2016financial}) and random perturbations could occasionally end up being above the curve,\footnote{This may also be linked to finite sample statistics and to the use of the optimal stationary strategy rather than the time-dependent one close to time $T$.} but our approach appears to be a very good one from a risk-reward perspective.\\

It must be noted that there is a significant difference between the efficient frontier of Markowitz modern portfolio theory and ours in that the expected P\&L is bounded from above in our case. Fig.~\ref{risk_reward} shows the maximum expected P\&L with no risk management (one still has to optimize quotes to make the most of the available flow). The simulated optimal curve saturates at a lower value even when $\gamma \to 0$ because of the inventory limit we imposed to build a grid-based finite difference scheme.\\

The choice of $\gamma$ ultimately rests with the dealer and it is clear that the optimal risk-reward curve can be useful in making the decision if one wants to optimize a risk-adjusted financial performance measure such as Sharpe ratio. FX dealers often have other  objectives than those purely based on risk-adjusted financial performance. For instance, they often care about market share. Although our model does not include such criterion, simulations similar to those carried out above could help choosing a value of $\gamma$, and thus strategies, that provide good results even when additional criteria are taken into account.\\

\section*{Concluding Remarks}

We have introduced and analyzed numerically a model of optimal market making incorporating fundamental risk controls: pricing ladders over a distribution of sizes and client tiers as well as the rate of hedging in external markets.
The model has immediate practical application to FX where the marketplace is significantly fragmented and dealers continuously solve the dilemma of whether to internalize or externalize their flow.
We have described relevant features of client flow taking a fictitious EURUSD franchise similar to that of a top-tier bank as an example and shown how tiers and pricing ladders as a function of size naturally arise from this analysis.
The results obtained regarding bid-ask spreads, risk neutralization times and internalization ratios are consistent with empirical data and publically reported figures.

\section*{Statement and acknowledgment}

The results presented in this paper are part of the research works carried out within the HSBC FX Research Initiative. The views expressed are those of the authors and do not necessarily reflect the views or the practices at HSBC. The authors are grateful to Richard Anthony (HSBC) and Paris Pennesi (HSBC) for helpful discussions and support throughout the project.

\bibliographystyle{plain}
\bibliography{mm_risk.bib}

\begin{thebibliography}{10}

\bibitem{almgren2001optimal}
Robert Almgren and Neil Chriss.
\newblock Optimal execution of portfolio transactions.
\newblock {\em Journal of Risk}, 3:5--40, 2001.

\bibitem{almgren2005optimal}
Robert Almgren, Chee Thum, Emmanuel Hauptmann, and Hong Li.
\newblock Direct estimation of equity market impact.
\newblock {\em Risk}, 18(7):58--62, 2005.

\bibitem{avellaneda2008high}
Marco Avellaneda and Sasha Stoikov.
\newblock High-frequency trading in a limit order book.
\newblock {\em Quantitative Finance}, 8(3):217--224, 2008.

\bibitem{bakshaev2020market}
Alexey Bakshaev.
\newblock Market-making with reinforcement-learning (sac).
\newblock {\em arXiv preprint arXiv:2008.12275}, 2020.

\bibitem{barzykin2021algorithmic}
Alexander Barzykin, Philippe Bergault, and Olivier Gu{\'e}ant.
\newblock Algorithmic market making in foreign exchange cash markets with
  hedging and market impact.
\newblock {\em arXiv preprint arXiv:2106.06974}, 2021.

\bibitem{bergault2021size}
Philippe Bergault and Olivier Gu{\'e}ant.
\newblock Size matters for otc market makers: general results and
  dimensionality reduction techniques.
\newblock {\em Mathematical Finance}, 31(1):279--322, 2021.

\bibitem{butz2019internalisation}
Maximilian Butz and Roel Oomen.
\newblock Internalisation by electronic fx spot dealers.
\newblock {\em Quantitative Finance}, 19(1):35--56, 2019.

\bibitem{cartea2015algorithmic}
{\'A}lvaro Cartea, Sebastian Jaimungal, and Jos{\'e} Penalva.
\newblock {\em Algorithmic and high-frequency trading}.
\newblock Cambridge University Press, 2015.

\bibitem{cartea2014buy}
{\'A}lvaro Cartea, Sebastian Jaimungal, and Jason Ricci.
\newblock Buy low, sell high: A high frequency trading perspective.
\newblock {\em SIAM Journal on Financial Mathematics}, 5(1):415--444, 2014.

\bibitem{cartea2019foreign}
\'{A}lvaro Cartea, Sebastian Jaimungal, and Jamie Walton.
\newblock Foreign exchange markets with last look.
\newblock {\em Mathematics and Financial Economics}, 13(1):1--30, 2019.

\bibitem{gueant2016financial}
Olivier Gu{\'e}ant.
\newblock {\em The Financial Mathematics of Market Liquidity: From optimal
  execution to market making}, volume~33.
\newblock CRC Press, 2016.

\bibitem{gueant2017optimal}
Olivier Gu{\'e}ant.
\newblock Optimal market making.
\newblock {\em Applied Mathematical Finance}, 24(2):112--154, 2017.

\bibitem{gueant2013dealing}
Olivier Gu{\'e}ant, Charles-Albert Lehalle, and Joaquin Fernandez-Tapia.
\newblock Dealing with the inventory risk: a solution to the market making
  problem.
\newblock {\em Mathematics and financial economics}, 7(4):477--507, 2013.

\bibitem{guilbaud2013optimal}
Fabien Guilbaud and Huyen Pham.
\newblock Optimal high-frequency trading with limit and market orders.
\newblock {\em Quantitative Finance}, 13(1):79--94, 2013.

\bibitem{ho1981optimal}
Thomas Ho and Hans~R. Stoll.
\newblock Optimal dealer pricing under transactions and return uncertainty.
\newblock {\em Journal of Financial economics}, 9(1):47--73, 1981.

\bibitem{oomen2017aggregator}
Roel Oomen.
\newblock Execution in an aggregator.
\newblock {\em Quantitative Finance}, 17(3):383--404, 2017.

\bibitem{oomen2017lastlook}
Roel Oomen.
\newblock Last look.
\newblock {\em Quantitative Finance}, 17(7):1057--1070, 2017.

\bibitem{ranaldo2021}
Angelo Ranaldo and Fabricius Somogyi.
\newblock Asymmetric information risk in fx markets.
\newblock {\em Journal of Financial Economics}, 140(2):391--411, 2021.

\bibitem{schrimpf2019fx}
Andreas Schrimpf and Vladyslav Sushko.
\newblock Fx trade execution: complex and highly fragmented.
\newblock {\em BIS Quarterly Review, December}, 2019.

\bibitem{schrimpf2019sizing}
Andreas Schrimpf and Vladyslav Sushko.
\newblock Sizing up global foreign exchange markets.
\newblock {\em BIS Quarterly Review, December}, 2019.

\end{thebibliography}

\end{document}